# Perfect Codes for Uniform Chains Poset Metrics

Tuvi Etzion, *Fellow, IEEE*

*Abstract*—The class of poset metrics is very large and contains some interesting families of metrics. A family of metrics, based on posets which are formed from disjoint chains which have the same size, is examined. A necessary and sufficient condition, for the existence of perfect single-error-correcting codes for such poset metrics, is proved.

*Index Terms*—Disjoint uniform chains, perfect codes, poset codes.

## I. Introduction

To be posted

### Acknowledgement

The author would like to thank Marcelo Firer for introducing the poset metrics for him.

T. Etzion is with the Department of Computer Science, Technion — Israel Institute of Technology, Haifa 32000, Israel. (email: etzion@cs.technion.ac.il).

This work was supported in part by the Israeli Science Foundation (ISF), Jerusalem, Israel, under Grant 230/08.

The work was partially done when the author visited the Centre Interfacultaire Bernoulli at EPFL, Lausanne, Switzerland in July 2011, supported by the Swiss National Science Foundation.